\documentclass[preprint,showpacs,aps]{revtex4}
\usepackage{graphicx}

\begin{document}

\title{A self-synchronizing stream cipher based on chaotic coupled maps}
\author{Shihong Wang$^{1}$, Gang Hu$^{2}$}
\email{ganghu@bnu.edu.cn} \affiliation{$^{1}$School of sciences, Beijing University of Posts and\\
Telecommunications,\\
Beijing 100876,China\\
$^{2}$Department of Physics, Beijing Normal University, Beijing 100875,\\
China}
\date{\today}

\begin{abstract}
A revised self-synchronizing stream cipher based on chaotic
coupled maps is proposed. This system adds input and output
functions aim to strengthen its security. The system performs
basic floating-point analytical computation on real numbers,
incorporating auxiliarily with algebraic operations on integer
numbers.
\end{abstract}


\maketitle In the recent tens years, since the pioneer work of
chaos synchronization by Pecora and Carrol \cite{r1}, secure
communication by utilizing chaos has attracted much attention.
Most of the secure communication systems suggested are based on
chaos synchronization and they
belong to self-synchronizing stream ciphers (SSSC)\cite%
{r2,r3,r4,r5,r6,r7,r8,r9,r10}. For SSSC the signals transmitted in
the public channels (i.e., ciphertext) serve as the drivings for
the synchronization of the receivers, and this structure can be
effectively used
by intruders to expose much information of the secret key\cite%
{r11,r12,r13,r14,r15,r16}. In \cite{r17,r18,r19,r20} we suggested
to use spatiotemporal chaos, i.e., one-way coupled chaotic map
lattices, incorporating with some simple conventional algebraic
operations, to enhance the security of chaotic SSSC. In this paper
we approach a revised system adding input and output functions aim
to strengthen the security of the system. The key length is 128
bits, and the key can be expanded to sub-keys to be used. In this
cipher all keys, plaintexts, ciphertexts and keystreams are
integers of 32 bits defined in $[0,2^{32})$.

\section{Encryption transformation}

Encryption transformation of the transmitter has two parts: the
keystream generator and encryption function. The keystream
generator utilizes one-way coupled logistic maps to produce the
keystream, incorporating with few simple algebraic operations on
integer numbers. The keystream generator has three parts: the
input function; coupled chaotic maps; and the output function.

The input function:
\begin{eqnarray}
D_{n+1} &=&E(FE(FE(FE(c_{n}\oplus k_{9})\oplus k_{10})\oplus
k_{11}\oplus
k_{12})  \nonumber \\
x_{n+1}(0) &=&\frac{D_{n+1}}{2^{32}-1}  \eqnum{1a}
\end{eqnarray}%
Where $c_{n}$ is the nth ciphertext and $x_{n}(0)$ is the state
variable for
the nth interaction of the 0th map. Symbol $\oplus $ is bitwise XOR. $%
k_{9},k_{10},k_{11},k_{12}$ are the expanded sub-keys to be defined later. $%
E $ and $F$ are nonlinear functions:

\begin{eqnarray}
F(A) &=&A\oplus (A>>>3)\oplus (A<<<11)  \nonumber \\
E(A) &=&A_{1}^{\prime }|A_{2}^{\prime }|A_{3}^{\prime
}|A_{4}^{\prime
}=s(A_{1})|s(A_{2})|s(A_{3})|s(A_{4})  \eqnum{1b} \\
A &=&A_{1}|A_{2}|A_{3}|A_{4}  \nonumber
\end{eqnarray}%
where the operation $A>>>(<<<)m$ denotes a right (left) cycle
shift of $A$ by $m$. Nonlinear function $E(A)$ is performed as
follows: first, we split a given 32-bit number $A$ into four bytes
$A_{1}|A_{2}|A_{3}|A_{4}$, according to the order from high to low
bit significance. Every byte takes a nonlinear map to
$A_{i}^{\prime }$, $i=1,2,3,4$(i.e., S-box operation) from
$A_{i}$, and then we obtain transformed four bytes $A_{1}^{\prime
}|A_{2}^{\prime }|A_{3}^{\prime }|A_{4}^{\prime }$. Finally, four
new bytes are combined to a new 32-bit integer $A^{\prime }$. The
input function guarantees that any one bit difference in the input
ciphertext $c_{n+1}$ can possibly change all 32 bits of the input
$D_{n+1}$.

The one-way coupled maps:%
\begin{eqnarray}
x_{n+1}(i) &=&\varepsilon f_{i}(x_{n}(i))+(1-\varepsilon
)f_{i}^{\prime
}(x_{n}(i-1))  \nonumber \\
f_{i}(x) &=&(3.75+\frac{a_{i}}{4})x(1-x),a_{i}\in \lbrack 0,1]  \nonumber \\
f_{i}^{\prime }(x) &=&(3.75+\frac{b_{i}}{4})x(1-x),b_{i}\in
\lbrack 0,1]
\eqnum{1c} \\
a_{i}
&=&\frac{k_{i}}{2^{32}-1},b_{i}=\frac{k_{i+4}}{2^{32}-1},i=1,2,3,4
\nonumber \\
y_{n+1} &=&int(x_{n+1}(4)\times 2^{52})%
\mathop{\rm mod}%
2^{32}  \nonumber
\end{eqnarray}%
where $\varepsilon $ is the coupled parameter, and $\varepsilon =2^{-16}$%
.The integer key $(k_{1},k_{2},k_{3},k_{4})$ and
$(k_{5},k_{6},k_{7},k_{8})$ are changed to real number $a_{i}$ and
$b_{i}$, $i=1,2,3,4$, for the computations of the logistic maps.

The output function:%
\begin{equation}
z_{n+1}=E(E(FE(Fy_{n+1}\oplus k_{13})\oplus Fy_{n+1}\oplus
k_{14})\oplus k_{15})  \eqnum{1d}
\end{equation}%
where $z_{n+1}$ is the keystream of the system.

The ciphertext $c_{n+1}$ is produced by%
\begin{equation}
c_{n+1}=z_{n+1}\oplus I_{n+1},%
\mathop{\rm mod}%
2^{32}  \eqnum{1f}
\end{equation}

where $I_{n+1}$ is the corresponding plaintext.

\section{The decryption transformation}

The decryption transformation of the receiver has also two parts,
the keystream generator and decryption function. The keystream
generator of the receiver is exactly the same as that of the
transmitter and the decryption function is written as

\begin{equation}
I_{n+1}^{\prime }=z_{n+1}\oplus c_{n+1},%
\mathop{\rm mod}%
2^{32}  \eqnum{2}
\end{equation}%
With the same key as that of the transmitter, the receiver can
reach chaos synchronization with the transmitter, and successfully
recover the true plaintext as

\begin{equation}
z_{n+1}^{\prime }=z_{n+1},I_{n+1}^{\prime }=I_{n+1}  \eqnum{3}
\end{equation}

\section{Key expansion}

The sub-keys in the system are derived from the main keys. With
the 128-bit main key $(k_{1},k_{2},k_{3},k_{4})$, the other
sub-keys are produced as

\begin{equation}
k_{i}=E(F(k_{i-4})\oplus k_{i-3}\oplus k_{i-2}\oplus
k_{i-1}),i=5,6,...,15 \eqnum{4}
\end{equation}

Every sub-key is defined in $[0,2^{32})$.

\section{S-box transformation}

S-box is a nonlinear transformation, and usually used in
cryptosystems. In our system s-box is defined by a transformation
from another 8-bit integer to 8-bit integer. We produce random
maps, $y_{i}=s(x_{i})$, and choose one form them with optimal
statistic properties. The input values of this map are
0,1,2,...,255, and the following output values are:

181 176 64 243 172 14 177 8 90 30 15 133 207 38 130 41

66 76 111 2 221 163 115 236 193 211 145 137 35 79 233 155

125 18 190 92 187 156 94 58 81 21 225 158 153 52 1 85

69 88 140 185 146 254 166 223 151 87 12 143 170 113 231 249

159 97 216 164 78 253 227 54 106 229 121 28 13 240 37 16

99 210 217 57 77 252 129 114 110 199 220 184 161 232 10 107

101 84 178 202 157 165 239 128 228 132 235 183 27 43 5 104

112 61 150 116 17 148 25 242 138 192 29 51 230 191 213 26

180 70 194 238 105 63 255 251 11 134 123 135 42 173 212 89

247 4 141 175 209 83 147 196 53 59 222 82 20 246 23 234

96 168 224 56 203 204 86 32 144 218 33 50 241 188 118 80

182 75 55 179 136 39 142 34 208 98 174 131 122 244 0 124

95 40 22 103 205 68 72 160 49 46 102 67 62 198 36 214

74 169 24 127 152 126 60 201 226 245 71 186 117 195 149 91

200 93 3 206 45 48 237 139 108 73 119 109 31 154 7 197

6 19 215 171 120 9 248 250 100 167 219 47 189 65 44 162.

\end{document}